\RequirePackage{fix-cm}
\documentclass[smallextended]{svjour3}       
\smartqed  
\usepackage{graphicx}
\usepackage{amssymb}
\begin{document}

\title{LRS Bianchi I model with constant deceleration parameter
}


\author{Vijay Singh         \and
        Aroonkumar Beesham 
}


\institute{\at
Department of Mathematical Sciences,\\
            University of Zululand,
              Private Bag X1001 \\
              Kwa-Dlangezwa, 3886,  \\
              South Africa \\
              \email{gtrcosmo@gmail.com (V. Singh) and beeshama@unizulu.ac.za (A. Beesham)}           
           }

\date{Received: date / Accepted: date}

\maketitle

\begin{abstract}
An LRS Bianchi I model is considered with constant deceleration parameter, $q=\alpha-1$, where $\alpha\geq0$ is a constant. The physical and kinematical behaviour of the models for $\alpha=0$ and $\alpha\neq0$ is studied in detail. The model with $\alpha=0$ describes late time acceleration, but eternal inflation demands a violation of the NEC and WEC. The acceleration is caused by phantom matter which approaches a cosmological constant at late times. The solutions with a scalar field also show that the model is compatible with a phantom field only. A comparison with the observational outcomes indicates that the universe has entered into the present accelerating phase in recent past somewhere between $0.2\lesssim z\lesssim0.5$. The model obeys the ``cosmic no hair conjecture". The models with $0<\alpha<1$ describe late time acceleration driven by quintessence dark energy. A violation of the NEC and WEC is required to accommodate the early inflationary epoch caused by phantom matter. The models with $1<\alpha<3$ describe decelerating phases which are usually occur in the presence of dust or radiation. These models are also found anisotropic at early times and attain isotropy at late times. The model for $\alpha=3$ represents a stiff matter era which also has shear at early stages and becomes shear free at late times, but it evolves with an insignificant ceaseless anisotropy. The models with $\alpha>3$ violate the DEC and the corresponding scalar field models have negative potential which is  physically unrealistic.

\end{abstract}

\section{Introduction}
\label{intro}
Present observations and theoretical schemes indicate that the evolution of the universe
comprises two periods of accelerated expansion, viz., the inflationary epoch \cite{GuthPRD1981,LindePLB1982,AlbrechtsSteinhardtPRL1982} and the present accelerating phase \cite{Riess1998,Perlmutteretal1999,Schmidtetal1998}. In both phases, the expansion is usually hypothesized as a result of the presence of matter with negative pressure dominating over the rest energy budget. This unknown matter  known as dark energy (DE) is responsible for the present accelerating phase, while it is dubbed primordial DE during the inflationary epoch. One of the main goals in theoretical and observational cosmology is to ascertain the nature of DE \cite{BahcalletalSc2001,JaffeellPRL2001}.

A cosmological constant $\Lambda$, which represents vacuum energy with equation of state (EoS) $\omega=-1$, is the simplest candidate for DE \cite{SahniStarobinsky2000,Padmanabhan2003,PeeblesRatra2003,Carroll2001}. The model based on a cosmological constant and cold dark matter (CDM) is well known as the  $\Lambda$CDM model. This model successfully explains the present acceleration of the universe, and fits well with observational data \cite{Adeetal2016,Komatsuetal2011}. However, the constraints on the EoS of DE from different observations \cite{BeanetalPRD2001,HannestadMorstellPRD2001,MelchiorrietalPRD2003} suggest an EoS for DE different from $\omega=-1$. The forms of DE different from $\Lambda$ are divided in two categories, namely, ``quintessence" \cite{ChibaPRD1999,Amendola2000PRD62} (see \cite{MartinMPLA2008} for detailed review) and ``phantom" \cite{CaldwellPLB2002,CaldwellPRL2003}, which are characterized by their respective EoS, $-1\leq\omega\leq-1/3$ and $\omega<-1$, respectively. Phantom matter is being regarded as one of the interesting possibilities that violates the null energy condition (NEC)\footnote{$\rho+p\geq0$}, whereas quintessence violates only the strong energy condition (SEC)\footnote{$\rho+3p\geq0$}. The existence of phantom matter is motivated by supernova data \cite{TonryetalApJ2003,Alametal0311364,ChoudhuryPadmanabhan0311622}, and the possibility that it might be currently dominating is supported by constraints on the  effective EoS \cite{KnopetalApJ2003,AlcanizPRD0401231}.

On the other hand, there is always a scope of following reverse approach in general relativity (GR), i.e., finding the matter required to give a desired geometric state. One can specify the scale factor(s), deceleration parameter, Hubble parameter, or a suitable relation between/among them to determine the effective equation of state of matter required to yield such behaviour. Following this procedure, Berman \cite{BermanNC1983} proposed a special law of variation for the Hubble parameter
\begin{equation}
  H=\beta a^{-\alpha},
\end{equation}
where $\alpha\geq0$ and $\beta>0$ are constants. Here $a(t)$ is the scale factor of the  Friedmann-Lemaitre-Robertson-Walker (FLRW) model. The deceleration parameter, $q=-1-\dot H/H^2$ (a dot denotes the ordinary derivative with respect to cosmic time $t$), for the above law returns the constant value
\begin{equation}
  q=\alpha-1.
\end{equation}
Therefore, Berman's law can also be referred as the law of constant deceleration parameter. The models with $\alpha<1$ correspond to inflating universes, whereas the models with $\alpha>1$ describe decelerating universes. A striking coincidence is that all the known cosmological models of Brans-Dicke theory with flat space-times naturally render a constant deceleration parameter \cite{JohriDesikanGRG1994}. \\
\indent The Hubble parameter is defined as $H=\dot a/a$, therefore, Eq. (1) yields \cite{BermanGomideGRG1988}
\begin{equation}
a(t)=\left\{
  \begin{array}{ll}
     a_1e^{\beta t}, & \hbox{$\alpha=0$;} \\
    \left[\alpha\left(\beta t+t_1\right)\right]^\frac{1}{\alpha}, & \hbox{$\alpha\neq0$,}
  \end{array}
\right.
\end{equation}
where $a_1$ and $t_1$ are integration constants. Hence, Berman's law is a generalization of the de Sitter and power-law expansions. The models with $\alpha\neq0$ have singular origin while those with $\alpha=0$ have non-singular origin.

Inserting Eq. (3) into Eq. (1), one has
\begin{equation}
H=\left\{
  \begin{array}{ll}
    \beta, & \hbox{$\alpha=0$;} \\
    \beta\left[\alpha(\beta t+t_1)\right]^{-1}, & \hbox{$\alpha\neq0$.}
  \end{array}
\right.
\end{equation}
Many authors have been presented solutions of Einstein's equations using Berman's law in different contexts of GR \cite{BermanGomideGRG1988,SinghSinghNC1991,BermanPRD1991,BermanGRG1991,BeeshamGRG1993,BeeshamPRD1993}. A number of authors have also been used this law in cosmological models in alternative and modified theories of gravity \cite{BermanGomideGRG1988,BermanSomIJTP1990,MaharajNaidooASS1993,JohriDesikanPJP1994,JohriDesikanALC1996,SinghDesikanPJP1997,PradhanetalIJMPD2001,SinghSinghASS2013}. All these referred works were carried out in homogenous and isotropic space-times.

Maharaj and Naidoo \cite{MaharajNaidooASS1993} suggested that the solutions presented in an isotropic background may be extended to the other models with less symmetry, and they also exemplified the case of the Bianchi I metric. The FLRW metric is a particular case of anisotropic space-times. Therefore, assuming Berman's law and defining the average scale factor in terms of directional evolutions, one can also find the solutions of the field equations in any anisotropic model. Pradhan and Vishwakarma \cite{PradhanVishwakarmaIJPAM2002} initiated this approach to obtain exact solutions of a locally-rotationally-symmetric (LRS) Bianchi I inhomogeneous model via a constant deceleration parameter, which  was later extended further in Lyra's geometry \cite{PradhanVishwakarmaJGP2004} by themselves. Pradhan and Yadav \cite{PradhanYadavIJMPD2002} considered a power-law expansion of the average scale factor to study a number of anisotropic models with bulk viscosity. Rahaman {\it et al} \cite{RahamanetalASS2005} also considered the power-law expansion to study an LRS Binachi I homogenous model in Lyra's geometry. Singh and Kumar implemented Berman's law to Bianchi II space-times with a perfect fluid \cite{SinghKumarIJMPD2006} and scalar field \cite{SinghKumarPJP2007}. Reddy {\it et al} considered a power-law to find solutions of LRS Bianchi I models in a scalar-tensor theory \cite{ReddyetalASS2006} and the scale-covariant theory \cite{ReddyetalASS2007}. Kumar and Singh \cite{KumarSinghASS2007} found exact solutions of Bianchi I models containing a perfect fluid assuming a constant expansion rate. Later on, many others have considered Berman's law to determine the solutions of Einstein's equations in various anisotropic space-time models. However, a very basic study of the LRS Bianchi I model with the consideration of Berman's law is still lacking. The properties of the effective matter for a constant deceleration parameter in LRS Bianchi I space-time needs to be explored in GR. The purpose of the present study is to fill this gap. We study an LRS Bianchi I space-time model for all possible values of $\alpha$ in (1)--(4), and determine the behavior of effective matter through the behavior of EoS and various energy conditions. We also explore the geometrical properties in each case. We extend our study to the case of a scalar field (normal or phantom) model.

The work is organised as follows. The field equations of the LRS Bianchi I model in Einstein's gravity are presented in Sec. 2. In Sec. 3, we present the solution of the field equations for $\alpha=0$ and explore the physical properties of the effective matter via an effective EoS under the constraints for a physically realistic scenario. In subsection 3.1, we discuss the possibility of eternal inflation for this model. The geometrical behavior of the model is discussed in subsection 3.2. In subsection 3.3, we consider the scalar field (quintessence or phantom) model. The solutions for $\alpha\neq0$ are obtained in Sec. 4 and its subsections. The same procedure is repeated to make the geometrical and physical interpretation of the solutions in this case what we follow for $\alpha=0$. The findings are summarized in Sec. 5.

\section{LRS Bianchi I model}
\label{sec:1}
The spatially homogenous and anisotropic LRS Bianchi I space-time metric is given as
\begin{equation}
ds^{2} =dt^{2}-A^2dx^2-B^2(dy^2+dz^2),
\end{equation}
where $A$ and $B$ are the scale factors, and are functions of the cosmic time $t$.\\
\indent The average scale factor and average Hubble parameter, respectively, are defined as
\begin{eqnarray}
  a&=&(AB^2)^{\frac{1}{3}},\\
  H&=&\frac{1}{3}\left(\frac{\dot A}{A}+2\frac{\dot B}{B}\right).
\end{eqnarray}
In GR, the Einstein field equations for the metric (5), with the consideration of a perfect fluid energy-momentum tensor, yield
\begin{eqnarray}
  \left(\frac{\dot B}{B}\right)^2+2\frac{\dot A\dot B}{A B}&=&\rho,\\
\left(\frac{\dot B}{B}\right)^2+2\frac{\ddot B}{ B}&=&-p,\\
\frac{\ddot A}{A}+\frac{\ddot B}{B}+\frac{\dot A \dot B}{AB}&=&-p,
\end{eqnarray}
where $\rho$ is the energy density and $p$ is the thermodynamical pressure of the effective matter.

From (9) and (10), the condition of isotropy of pressure is
\begin{equation}
\frac{\dot A}{A} -\frac{\dot B}{B}=\frac{k}{AB^2},
\end{equation}
\noindent where $k$ is a constant of integration.

By the use of (4), one can easily solve Eqs. (7) and (11). In what follows we shall present the solutions for $\alpha=0$ and $\alpha\neq0$, and study the physical and geometrical properties of the both models.

\section{Model with $\alpha=0$}
\label{sec:2}
For $\alpha=0$, we have $H=\beta$, and $a^3=AB^2=(a_1e^{\beta t})^3$, for which (7) and (11) admit the solutions
\begin{eqnarray}
  A&=&c_1e^{\beta  t-\frac{2 k }{9  \beta }e^{-3 \beta  t}},\\
B&=&c_1e^{\beta  t+\frac{k }{9  \beta }e^{-3 \beta  t}},
\end{eqnarray}
where $c_1$ is an integration constant and $a_1$ is taken to be unity without loss of generality.

The average scale factor in terms of red shift $z$ is defined via the relation $a(t)=a_0(1+z)^{-1}$, where $a_0$ is the present value of scale factor. This relation is used to express physical quantities and parameters in terms of $z$. The energy density and pressure become
\begin{eqnarray}
  \rho&=&3 \beta ^2-\frac{1}{3} k^2 e^{-6 \beta  t}=3 \beta ^2-\frac{1}{3}k^2(1+z)^6,\\
  p&=&-3 \beta ^2-\frac{1}{3} k^2 e^{-6 \beta  t}=-3 \beta ^2-\frac{1}{3}k^2(1+z)^6,
\end{eqnarray}
where we have taken $a_0=1$.

The variable term in (14) and (15) decreases with time, consequently, the energy density and pressure increase with the cosmic evolution and approach  the constant value $\rho=3\beta^2=-p$ as $t\to\infty$. The directional scale factors (12) and (13) show that the model has an infinite past but  $\rho\to-\infty$ as $t\to-\infty$, which makes the model physically unrealistic. For a physically realistic scenario the energy density must be positive. For $\rho\geq0$, we must have
\begin{equation}
  t\geq\frac{1}{\beta}\ln\left(\frac{k^2}{9\beta^2}\right)^\frac{1}{6},\;or\;
  z\leq\left(\frac{9\beta^2}{k^2}\right)^\frac{1}{6}-1.
\end{equation}
It is natural because we are demanding an accelerating expansion ($q=-1$) from an anisotropic model, so the above constraint implies that the LRS Bianchi I model can exhibit acceleration only when the universe enters into an accelerating phase.

The EoS parameter, which is defined as $\omega=p/\rho$, yields
\begin{equation}
  \omega=\left[-1+\frac{2 }{1+\gamma e^{6 \beta t}}\right]^{-1}=\left[-1+\frac{2 }{1+\gamma (1+z)^{-6}}\right]^{-1}.
\end{equation}
where $\gamma=9\beta^2/k^2$. The above expressions of the EoS in $t$ and $z$ coincide when $t=0$ and $z=0$, due to the consideration of the present scale factor to be unity. As we know $z=0$, corresponds the present universe, therefore, $t=0$, represents the present time in this model. The present value of the EoS parameter $\omega_0(z=0)$, is
\begin{equation}
  \omega_0=\frac{1+\gamma}{1-\gamma}.
\end{equation}
From this expression we have $\omega_0>1$ for $0<\gamma<1$ and $\omega_0<-1$ for $\gamma>1$. Since the redshift $z=\left(9\beta^2/k^2\right)^{1/6}-1$ or $z=\gamma^{1/6}-1$, represents a past event (when the universe started accelerating), therefore, it must be positive, which is possible when $\gamma>1$. Hence, the model shows that the universe at present is dominated by phantom DE. Let us probe this theoretical outcome with the observational data.

Combined results from the Cosmic Microwave Background (CMB) experiments with Large Scale Structure (LSS) data, the $H(z)$ measurement from the Hubble Space Telescope (HST), and luminosity measurements of Type Ia Supernovae (SNe Ia), put the constraints on the EoS, $-2.68 <\omega_{obs}< -0.78$ \cite{MelchiorrietalPRD2003}. These bounds become even more tight, i.e., $-1.45 <\omega_{obs}< -0.74$ \cite{HannestadMorstellPRD2001}, when the WMAP data (also see refs. \cite{Alametal0311364,AlcanizPRD0401231}) is included. For these observational limits, from (18), we get $\gamma>2.19$, for the former bounds and $\gamma>5.44$, for the latter. This observational analysis confirms that the theoretical prediction is correct. Now we sketch out the finite past, for the redshift given in (16), and future dynamics of the EoS parameter (17) for some values satisfying these constraints.

\begin{figure}[h]
\includegraphics{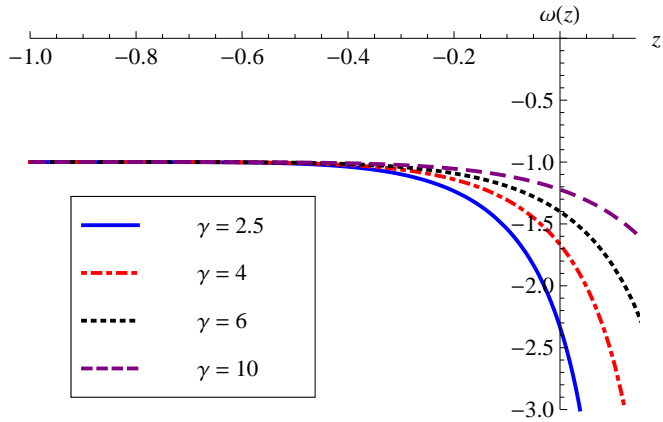}
  \caption{$\omega$ {\it \emph{\emph{versus}}} $z$ for different values of $\gamma$.}
  \label{fig:1}
\end{figure}

Figure 1 plots the behavior of the EoS parameter against redshift $z<\gamma^{1/6}-1$  for some values of $\gamma>2.19$. We see that $\omega$ diverges at $z=\gamma^{1/6}-1$ due to the vanishing of the energy density at this point. Therefore, although it is not a good idea to use the EoS (17) to depict the past behavior of the effective matter, it can be used to determine the present and future behavior. Fig. 1 clearly shows that the present acceleration of the universe is caused by the influence of the effective phantom matter. The behavior of the phantom matter asymptotically approaches a cosmological constant in future. In Sec. 3.3, via a consideration of a scalar field as effective matter, we shall show that this model is compatible only with a phantom scalar field. According to the observational limits of $\omega_0$ mentioned before, our model indicates that the universe has entered into the present accelerating phase very recently,  somewhere between $0.2\lesssim z\lesssim0.5$.

Thus, if we demand cosmic acceleration from an anisotropic model,  the solutions become valid only after a period when the universe transits into an accelerating phase. However, if one agrees to consider the matter energy density to be negative before the time given in (16), then the solutions for $\alpha=0$ describe an eternal inflating universe. Let us discuss this phenomenon in detail.

\subsection{Eternal inflation}
\label{sec:3}
Although phantom matter is theorised usually to describe late time cosmic acceleration (see \cite{NojiriOdintsovGRG2006} for an extensive list of references on these studies), phantom cosmology allows us to account for the dynamics and matter content of the universe tracing back the evolution to the inflationary epoch. If phantom energy is currently dominating over all other cosmic components in the universe, then we were starting a period of super-inflation which would be accelerating more rapidly than de Sitter expansion. There are possibilities that inflation may have been driven by a matter component that violates the NEC, thereby leads to super-inflation which has already been considered by many researchers \cite{PiaoZhouPRD2003,DiazMadridPLB2004,PiaoZhangPRD2004,AnisimovetalJCAP2005,BaldietalPRD2005}.

Most inflationary cosmological models are described by  homogeneous and isotropic cosmology. The existence of analytical solutions of the field equations in most of the cases justifies this oversimplification of the geometry of space-time. In flat FLRW cosmology, inflation refers to de Sitter expansion driven by a positive cosmological constant, where the scale factor grows exponentially. However, an exponential expansion is by no means uniquely required for acceleration. Rather, the violation of the SEC is what uniquely characterizes accelerated expansion. With this enlarged definition, many solutions of the Einstein equations, different from the de Sitter one, may describe an inflating universe, for example, a power-law expansion ($a(t)\propto t^n$, $n>1$) also works well \cite{AbbottWiseNPB1984}. In fact, inflation can be given the broader meaning as to correspond to expansion that accelerates. This holds true in particular for spatially homogeneous and non-isotropic universes as there are no compelling physical reasons to assume homogeneity and isotropy before the inflationary period. Consequently, a number of non-FLRW models have been investigated to determine whether an inflationary epoch occurs in an anisotropic background.

A preliminary investigation of compatibility of inflation and anisotropy was undertaken by Barrow and Turner \cite{BarrowTurnerNature1981}. They pointed out that in the original inflationary mechanism proposed by Guth \cite{GuthPRD1981}, large anisotropy prevents the de Sitter phase from occurring, however, a universe with moderate anisotropy will undergo inflation. On the other hand, Demianski \cite{DemianskiNature1984} showed that the new inflationary scenario \cite{LindePLB1982,AlbrechtsSteinhardtPRL1982} is compatible with large initial anisotropy. Steigman and Turner \cite{SteigmanTurnerPLB1983} also came to the same result that neither shear nor negative curvature prevents inflation. Wald \cite{WaldPRD1983} independently investigated the asymptotic behavior of initially expanding homogeneous cosmological models. He concluded that almost all homogeneous and anisotropic models (except Binachi IX)\footnote{The behavior of the Bianchi IX model is similar, provided that the  cosmological constant must be large compared to the spatial curvature.}, undergo inflation, exponentially evolving towards an isotropic de Sitter solution.

In the present model, we have
\begin{equation}
  \rho+3p=-6 \beta ^2-\frac{4 }{3 }k^2 e^{-6 \beta  t},
\end{equation}
which remains negative throughout the cosmic evolution, and hence insures that the SEC is always violated. This is not however an outcome of the study, but a consequence of an adhoc assumption, $q=-1$ (ever inflating model). Since the effective matter has negative energy density before the cosmic time calculated in (16), if a violation of the weak energy condition\footnote{$\rho\geq0$, $\rho+p\geq0$} (WEC) is allowed, the model describes eternal inflation.

Rothman and Madsen \cite{RothmanMadsenPLB1985} examined the arguments of inflation in Bianchi I anisotropic cosmologies, and found that they contain serious inconsistencies. However, the authors also presented several alternatives to remedy a few of these inconsistencies. They suggested that a proper treatment of Bianchi I inflation evidently requires a quantum theory of gravity. Although the energy density of a field in classical physics is strictly positive, the energy density in quantum field theory can be negative due to quantum coherence effects \cite{EpsteinetalNC1965}. The Casimir effect \cite{LamoreauxPRL1997,MohideenRoyPRL1998} and squeezed states of light \cite{WuetalPRL1986} are two familiar examples which have been studied experimentally. As a result, all the known pointwise energy conditions in GR, such as the WEC and NEC, are allowed to be violated. Even for a scalar field in  flat Minkowski spacetime, it can be proved that the existence of quantum states with negative energy density is inevitable \cite{EpsteinetalNC1965}. Our interpretation of eternal inflation with the violation of the WEC can be advocated by any  such mechanism. However, it is obvious to wonder whether a single fluid may indeed be credit-worthy for both periods of accelerating expansion. But there are studies \cite{NojiriOdintsovGRG2006,NojiriOdintsov0506212,Capozzielloetal0507182} to justify where phantoms are essential to unify inflation and late time acceleration of the universe. The behavior of the effective matter in the present model is also consistent to describe an eternal inflating universe.

\subsection{Geometrical behavior of the model}
\label{sec:4}
In general, a homogeneous and non-isotropic model may or may not inflate, but if it inflates, it may or may not have enough time to approach the de Sitter stage. The issue of isotropization in anisotropic inflationary cosmology is related to the ``cosmic no hair conjecture"\footnote{The conjecture states that a large class of ever-expanding cosmological models with a positive cosmological constant rapidly approach asymptotically  a de Sitter state at late times.} \cite{GibbonsHawkingPRD1977,HawkingMossPLB1982}. Many attempts have been made \cite{SteigmanTurnerPLB1983,GonzalezJonesPLB1986,TurnerWidrowPRL1986,MossSahniPLB1986} but this conjecture remains to be proven. Jensen and Stein-Schables \cite{JensenStein-SchablesPRD1986} generalized these attempts to analyze the effects of anisotropic cosmologies on inflation, where either the anisotropies were treated as a small perturbation on an FLRW background, or the anisotropic models contained an FLRW model as a special case. They showed that any Bianchi model (except maybe Bianchi IX)\footnote{If it is possible to enter into the inflationary phase well before the model recollapses, then all arguments still hold true for Bianchi IX also. However, this requires a comparison of two time scales that are strongly dependent on initial conditions.} that undergoes sufficient inflation will become isotropic on scales greater than the horizon today to a very high degree of accuracy, independent of the initial conditions.

From (12) and (13), we see $A=B=c_1e^{\beta t}=a(t)$ as $t\to\infty$, where we have chosen $c_1=a_1$, which shows that the model approaches  an isotropic de Sitter expansion at late times. Hence, the model obeys the ``cosmic no hair conjecture" as well. However, it is to be noted that the models that obey this conjecture are usually considered with a positive cosmological constant, but the present study evidences that this property can be accomplished without a cosmological constant, which is an interesting feature of the model. The same dynamics can be seen from the geometrical parameters.

The rates of the expansion along the $x$, $y$, and $z -$axes are
\begin{eqnarray}
  H_x=\frac{\dot A}{A}&=&\beta+\frac{2}{3} k  e^{-3 \beta t},\\
  H_y=H_z=\frac{\dot B}{B}&=& \beta-\frac{1}{3} k  e^{-3 \beta t}.
\end{eqnarray}
The expansion scalar, $\theta$ and the shear scalar, $\sigma$ are
\begin{equation}
 \theta=\frac{\dot A}{A}+2\frac{\dot B}{B}=3\beta,
\end{equation}
\begin{equation}
\sigma^2=\frac{1}{2}\sigma_{\mu\nu}\sigma^{\mu\nu}=\frac{1}{3}\left(\frac{\dot A}{A}-\frac{\dot B}{B}\right)^2=\frac{1}{3} k^2 e^{-6 \beta  t},
\end{equation}
\noindent where $\sigma_{\mu\nu}$ is the shear tensor. The universe expands uniformly, and the early evolution has shear but late time evolution becomes shear free.

The anisotropy parameter is
\begin{equation}
  \mathcal{A}=\frac{1}{3}\sum_{i=1}^3\left(\frac{H_i-H}{H}\right)^2=\frac{2 k^2 e^{-6 \beta  t}}{9 \beta ^2}=\frac{2 (1+z)^6}{\gamma},
\end{equation}
where $H_i$ ($i=1,2,3$) represents the directional Hubble parameters in the directions $x$, $y$, $z$ respectively. The anisotropic parameter shows that the universe was highly anisotropic in the infinite past,  but becomes isotropic at late times. The present value of the anisotropic parameter is $\mathcal{A}_0(z=0)=2/\gamma$, which shows that the present universe is negligibly anisotropic since at present $\gamma\geq2.19$ or $\gamma\geq5.44$, according to various observational data as discussed in Sec. 3.

Barrow \cite{BarrowPLB1987} showed that Bianchi I homogenous inflationary models with hypersurface-orthogonal velocity fields which possesses shear containing a perfect fluid with EoS, $p=\omega\rho$ approach  isotropy only with stiff matter. In contrast, the present model approaches  isotropy in the presence of phantom matter. This is not, in fact, a contradiction, since we have not used an  EoS to obtain the solution. Instead, we have followed a reverse approach, i.e., assuming the average scale factor for the LRS Bianchi I model evolving to de Sitter expansion, we have investigated the EoS which can give the desired geometrical evolution. We see that  $\omega\to1$ as $t\to-\infty$ or $z\to\infty$, which shows that the matter in the universe had the characteristics similar to stiff matter in the infinite past. However, this stiff matter is different from the usual stiff matter for which the WEC holds. The phantom matter in the present study acts as stiff matter in the infinite past but violates the WEC. Therefore, let us call it ``phantom stiff matter". We understand that there is no known physical motivation for the existence of such matter. However, on the theoretical side, the issue is now to incorporate a matter source which can act as phantom matter at present, and as a cosmological constant in future, and phantom stiff matter in the infinite past. The key for such matter is within the outcome of the present model, i.e., it is hypothetical phantom matter itself.

Phantom matter can originate from scalar fields with a global minimum in their effective potential \cite{LindeRPP1985}, from higher order curvature terms in higher-order theories of gravity \cite{BarrowOttewilJPA1983,HawkingLuttrellNPB1984,WhitePLB1984,MijicetalPRD1986,PollockPLB1988}, from a bulk viscous stress due to particle production \cite{BarrowNPB1988}, in Brans-Dicke and non-minimally coupled scalar field theories \cite{TorresPRD2002}, in strange effective quantum field theory \cite{NojiriOdintsovPLB5622003,NojiriOdintsovPLB5712003}, and by some other means (for detail see \cite{PiaoZhangPRD2004}). Most of these disparate prescriptions require some weakly coupled scalar field to be displaced from its equilibrium state. If the subsequent evolution towards the new equilibrium state proceeds sufficiently slowly, then inflation will result. Therefore, in what follows we shall consider a minimally coupled scalar field (normal or phantom) with scalar potential as the effective matter source in this model.

\subsection{Scalar field model}
\label{sec:5}
The energy density and pressure of a minimally coupled normal ($\epsilon=1$) or  phantom ($\epsilon=-1$) scalar field, $\phi$ with self-interacting potential, $V(\phi)$, are given by
\begin{eqnarray}
   \rho&=&\frac{1}{2}\epsilon\dot\phi^2+V(\phi),\\
   p&=&\frac{1}{2}\epsilon\dot\phi^2-V(\phi).
\end{eqnarray}
Substituting (25) and (26) in (14) and (15), we get the expressions for the kinetic energy and scalar potential
\begin{eqnarray}
  \frac{1}{2}\epsilon\dot\phi^2&=&-\frac{1}{3} k^2 e^{-6 \beta  t},\\
V(t)&=&3\beta^2,
\end{eqnarray}
respectively. The kinetic energy must be positive for reality of the solutions which is possible for $\epsilon=-1$. Hence, the model is compatible with a phantom scalar field only. The kinetic energy evolves with an infinite value which decreases with the evolution,  and vanishes at late times, while the potential remains flat throughout the evolution.

On integrating (27), we find the expression for the phantom scalar field
\begin{equation}
  \phi=\phi_0\pm\left(\frac{2}{3}\right)^\frac{1}{2}\frac{ke^{-3 \beta  t}}{3\beta}=\phi_0\pm\left(\frac{2}{3\gamma}\right)^\frac{1}{2}(1+z)^{3},
\end{equation}
where only a positive sign is compatible for physical consistency.

\begin{figure}
  \includegraphics{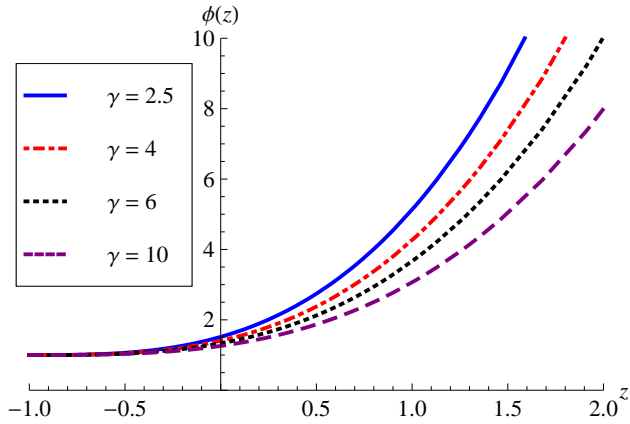}
\caption{$\phi(z)$ {\it \emph{\emph{versus}}} $z$ for different values of $\gamma$ with $\phi_0=1$.}
\label{fig:2}       
\end{figure}

The evolution of the scalar field with $z$ for some values of $\gamma$ consistent with observations is shown in Fig. 2. The scalar field decreases and approaches a finite minimum value at late times. The flat potential can be identified as a cosmological constant. The key difference between phantom (or quintessence) and a cosmological constant is that the energy density of the former could be a significant fraction of the overall energy even in the early universe, while the cosmological constant is dynamically relevant only at late times. One may readily verify that standard de Sitter solutions for a flat FRW model are recovered when $k=0$. If $k=0$ then $\phi=\phi_0$ and $\rho=3\beta^2=-p$, which essentially corresponds to a cosmological constant and the standard de Sitter model.

\section{Model with $\alpha\neq0$}

When $\alpha\neq0$, we have $H=\beta\left[\alpha(\beta t+t_1)\right]^{-1}$, and $a^3=AB^2=\left[\alpha\left(\beta t+t_1\right)\right]^{3/\alpha}$. Substituting these into (7) and (11), we obtain
\begin{eqnarray}
  A&=&\left\{
  \begin{array}{ll}
     c_2t^{\frac{1}{3}+\frac{2k }{9 \beta}}, & \hbox{$\alpha=3$;} \\
    c_2t^\frac{1}{\alpha} e^{\frac{2 k  \alpha t} {3 (\alpha-3)}(\beta \alpha t)^{-\frac{3}{\alpha}}}, & \hbox{$\alpha\neq3$,}
  \end{array}
\right.\\
B&=&\left\{
  \begin{array}{ll}
     c_2t^{\frac{1}{3}-\frac{k }{9 \beta}}, & \hbox{$\alpha=3$;} \\
    c_2t^\frac{1}{\alpha} e^{\frac{k  \alpha t} {9-3 \alpha}(\beta \alpha t)^{-\frac{3}{\alpha}}}, & \hbox{$\alpha\neq3$,}
  \end{array}
\right.
\end{eqnarray}
where $c_2$ is an integration constant and we have taken $t_1=0$, to shift the origin to  $t=0$.

\subsection{Model with $\alpha\neq3$}

For $\alpha\neq3$, the energy density and pressure become
\begin{eqnarray}
 \rho&=& \frac{3}{\alpha ^2 t^2}-\frac{1}{3} k^2 (\alpha  \beta  t)^{-\frac{6}{\alpha }},\\
p&=&\frac{2 \alpha -3}{\alpha ^2 t^2}-\frac{1}{3} k^2 (\alpha  \beta  t)^{-\frac{6}{\alpha }}.
\end{eqnarray}
We study the physical behavior of the model through energy conditions for which we need
\begin{eqnarray}
  \rho+3p&=&\frac{6 (\alpha -1)}{\alpha ^2 t^2}-\frac{4}{3} k^2 (\alpha  \beta  t)^{-\frac{6}{\alpha }},\\
  \rho+p&=&\frac{2}{\alpha  t^2}-\frac{2}{3} k^2 (\alpha  \beta  t)^{-\frac{6}{\alpha }},\\
\rho-p&=&\frac{2(3- \alpha) }{\alpha ^2 t^2}.
\end{eqnarray}
When $\alpha<3$, the energy density will be positive after a certain time, say $t=t'$. Similarly, when $\alpha>3$, the energy density will be positive before $t=t'$. However, due to the complicated expression of energy density it is not possible to find the explicit expression for such specific time where the energy density transits from positive to negative or vice-verse. The same is true for the expressions (34) and (35). Therefore, to study the energy conditions we shall use the terms `early times' for $t<t'$ and `late times' for $t>t'$. Let us divide our study into different intervals of $\alpha$, inflating model ($0<\alpha<1$) and decelerating model ($\alpha>1$).

\subsubsection{When $0<\alpha<1$}

The violation of the SEC for an ever inflating model is obvious, which can be seen from (34) as $\rho+3p$ remains always negative for $0<\alpha<1$. We note that the NEC and WEC are violated at early times, but hold at late times for arbitrary values of $\beta$ and $k$. Therefore, the model in general exhibits eternal inflation, however, if we respect the  NEC and WEC, then it is suitable to describe late time acceleration only. Thus, the physical behavior of the model in this case is similar to an eternal inflationary model as studied for $\alpha=0$. Hence, the discussion made in subsection 3.1 is true for this model too.

\subsubsection{When $1<\alpha<3$}

In this case the model describes a decelerated phase of the universe. The SEC is violated  at early times and holds at late times, but the violation of the SEC contradicts  decelerated expansion. Therefore, this model can be valid only at late times when inflation ends and the universe transits into a decelerating phase. The NEC and WEC also hold good at late times.

\subsubsection{When $\alpha>3$}

The model in this case also can represent decelerated universes. The NEC and WEC are obeyed at early stages of evolution, but the dominant energy condition\footnote{$\rho\geq|p|$, i.e., $\rho\pm p\geq0$} (DEC) is always violated, which is not a realistic cosmological scenario.

\subsection{Scalar field model}

From (25), (26), (32) and (33), the kinetic energy is
\begin{equation}
  \frac{1}{2}\epsilon\dot\phi^2=\frac{1}{\alpha  t^2}-\frac{1}{3} k^2 (\alpha  \beta  t)^{-\frac{6}{\alpha }},
\end{equation}
and the potential for the scalar field is
\begin{equation}
  V(t)=\frac{3- \alpha }{\alpha ^2 t^2}.
\end{equation}
The above scalar potential can play a significant role during early evolution, but vanishes at late time times. Let us divide our further discussion into the following cases for different intervals of $\alpha$.

\subsubsection{When $0<\alpha<1$}

The WEC for arbitrary values of $\beta$ and $k$, is obeyed only at late times. The kinetic energy at late times can be positive with a quintessence field only. Therefore, the model describes late time accelerating expansion in the presence of quintessence DE. However, if one allows to the WEC to be violated, then the solutions can be compatible with a phantom scalar field during early evolution, and the model accounts for the  inflationary period as well. Hence, if we respect the NEC and WEC, then the model describes only  late time acceleration, while it exhibits eternal inflation otherwise.

\subsubsection{When $1<\alpha<3$}

The model corresponds to decelerated universes in this case. The WEC holds at late times. The kinetic energy at late times can be positive with a quintessence scalar field. Therefore, decelerated phases occur in the presence of dust and/or radiation. If we allow the WEC to be violated at early times, then the solutions can also be compatible with a  phantom scalar field, but it would be irrelevant because decelerated expansion cannot occur in the presence of exotic matter.

\subsubsection{When $\alpha>3$}

The potential becomes negative in this case which is unrealistic. Nevertheless, if we agree to consider a negative potential\footnote{Quantum theory and phantom scalar fields allow a negative potential in some contexts (see for example \cite{GrahamOlumPRD2003} and references therein).} which is equivalent to allowing a violation of the DEC, then the model can be suitable to describe early evolution because the WEC is obeyed at early times in this case. The solutions during early evolution can be compatible with a normal scalar field only. Thus, the model can describe decelerated universes in the presence of dust and/or radiation. If we allow a violation of the WEC at late times, then the solutions can also be consistent with a phantom field, but they lead to a contradiction for a decelerated universe.

\subsection{Geometrical behavior}

The rates of the expansion along with the  $x$, $y$, and $z -$axes give
\begin{eqnarray}
  H_x=\frac{1}{\alpha  t}+\frac{2}{3} k (\alpha  \beta  t)^{-\frac{3}{\alpha }},\\
  H_y=H_z=\frac{1}{\alpha  t}-\frac{1}{3} k (\alpha  \beta  t)^{-\frac{3}{\alpha }}.
\end{eqnarray}
The expansion scalar, $\theta$ and the shear scalar, $\sigma$ become
\begin{eqnarray}
 \theta&=&\frac{3}{\alpha  t},\\
\sigma^2&=&\frac{1}{3} k^2 (\alpha  \beta  t)^{-\frac{6}{\alpha }}.
\end{eqnarray}
The expansion scalar and shear scalar decrease with the evolution, and vanish at late times. Hence, the early universe had shear but late universe becomes shear free.

The anisotropic parameter of the expansion is
\begin{equation}
  \mathcal{A}=\frac{2}{9} \alpha ^2 k^2 t^2 (\alpha  \beta  t)^{-\frac{6}{\alpha }}.
\end{equation}
The behavior of $\mathcal{A}$ depends on the values of $\alpha$. It decreases for $\alpha<3$, while increases for $\alpha>3$, with the evolution of the universe. Hence, the models with $\alpha<3$ were anisotropic at early times and asymptotically become isotopic at late times. On the other hand, the models with $\alpha>3$ behave in a vice-versa manner, which is not consistent with the observed universe. Hence, the models with $\alpha>3$ are not feasible geometrically too. Thus, the dynamics of these geometrical parameters is in accordance with the physical behaviors discussed in Secs. 4.1 and 4.2 and in subsections therein. The solutions for an isotropic model can be recovered by substituting $k=0$.

\subsection{Model with $\alpha=3$}

The model describes a decelerating universe in this case. The energy density and pressure become equal, i.e.,
\begin{eqnarray}
  \rho=  p=\frac{9 \beta^2- k^2}{27 \beta^2 t^2}.
\end{eqnarray}
We must have $9\beta^2\geq k^2$ for a physically realistic scenario. The solutions in this particular case represent the stiff matter phase of the universe.

\subsubsection{Geometrical behavior}

The rates of the expansion along the  $x$, $y$, and $z -$axes give
\begin{eqnarray}
  H_x&=&\left(\frac{1}{3}+\frac{2 k }{9 \beta}\right)\frac{1}{t},\\
  H_y=H_z&=&\left(\frac{1}{3}-\frac{ k }{9 \beta}\right)\frac{1}{t}.
\end{eqnarray}
The expansion scalar, $\theta$ and the shear scalar, $\sigma$ become
\begin{eqnarray}
 \theta&=&\frac{1}{t},\\
\sigma^2&=&\frac{k^2}{27 \beta ^2 t^2}.
\end{eqnarray}
The expansion scalar diminishes with the evolution of the universe. Similarly, the universe has shear at early times which dies out at late times.

The anisotropic parameter of the expansion is
\begin{equation}
  \mathcal{A}=\frac{2 k ^2}{9 \beta^2 }.
\end{equation}
The WEC holds for $k^2\leq9\beta^2$, therefore, we have $0<\mathcal{A}\leq2$. Hence, the stiff matter phase evolves with negligible changeless anisotropy. When $k=0$ the solutions reduce to an isotropic model.

These are known solutions to Einstein's equations with the properties (i) the ratio of the energy density to the square of the fluid's volume expansion is constant ($\rho\propto\theta^2$), and (ii) the ratio of the rate of shear of the fluid to the volume expansion is constant ($\sigma^2\propto\theta^2$). These properties are exhibited in certain solutions in the cases of Bianchi models \cite{JacobsAJ1968,CollinsCMP1971,EllisMacCallum1969}. Collins \cite{CollinsPL1977} showed that property (ii) is in fact a consequence of property (i), i.e., $\rho\propto\theta^2\Rightarrow\sigma^2\propto\theta^2$ whenever the matter content is a perfect fluid with a barotropic equation moving along an expanding geodesic and hypersurface-orthogonal time-like congruence, regardless of whether or not the space-time is spatially homogenous.

\section{Conclusion}
\label{sec:6}
We have studied LRS Bianchi I models with constant deceleration parameter, $q=\alpha-1$. We have explored the physical and geometrical properties of the models when $\alpha=0$ and $\alpha\neq0$. Knowing the EoS of DE is one of the biggest challenges in theoretical physics \cite{Linder2003,Rayetal2007,Mukhopadhyayetal2007,Linder2008,Rayetal2011} as well as in observational cosmology today \cite{Knopetal2003,Tegmarketal2004}. The observations seem to favor an evolving EoS for DE \cite{HutererTurner2001,HutererCooray2005}. The model with $\alpha=0$ leads to an evolving EoS for DE which is consistent with  observations. If we respect the WEC, then the model describes late time acceleration, but eternal inflation demands a violation of the WEC during early times. The present acceleration is caused by the influence of effective phantom matter which behaves like a cosmological constant as $t\to\infty$. It is observed that $\omega\to1$ as $t\to-\infty$, which shows that there might be a matter in the infinite past which has the characteristics similar to stiff matter, but as this matter violates the WEC, we call it ``phantom stiff matter". We have compared our results with the recent observational data and found that the universe has entered into present accelerating phase very recently somewhere between $0.2\lesssim z\lesssim0.5$.

An important characteristic of this model is, while the de Sitter expansion ($q=-1$) in a flat FLRW space-time is driven by a cosmological constant (vacuum energy), the acceleration in this model is driven by a dynamical DE. This is a striking difference of the LRS Bianchi I model from that of a flat FLRW model. This feature proves that anisotropic models are a prominent way to study the evolution of the universe. Another important aspect of this model is, while  phantom matter is considered an adhoc assumption as a candidate of DE in most of the forward approaches, its presence in an anisotropic model is a natural outcome in the reverse approach followed in this study.

Respecting the phantom behavior of the effective matter in this model, we have extended our solutions to the case of a scalar field (quintessence or phantom) model. The scalar field model also confirms that the solutions are consistent with a phantom scalar field only. We have obtained the scalar field and the scalar potential. The scalar field decreases with the evolution, while the scalar potential remains flat throughout. The flat potential can be regarded as a cosmological constant.

However, it is not necessary that the source of phantom matter could only be a scalar field. There are a number of mechanisms to induce the phantom kind of behavior of the matter. Nojiri and Odintsov \cite{NojiriOdintsovPLB5622003,NojiriOdintsovPLB5712003} considered a de Sitter and Nariai universe induced by quantum CFT (conformal field theory) with classical phantom matter and a perfect fluid. The model represents the combination of trace-anomaly driven inflation and a phantom driven de Sitter universe. The similarity of phantom matter with quantum CFT indicates that a phantom scalar field may be an effective description for some quantum field theory. Since the phantom matter in the present study violates the WEC in the early universe, one may look at this phenomenon in view of quantum field theory.

The model has been found highly anisotropic during early evolution, but the anisotropy is smoothed out at late times. More interestingly, the model obeys the ``cosmic no hair conjecture" and approaches the de Sitter state at late times. The universe at present has been found negligibly anisotropic. An inflating phantom universe seems to adamantly ends up in a catastrophic big-rip singularity. The model with $\alpha=0$, however, is free from any of such singularities.

It is to be noted that Gron \cite{GronPRD1985} derived a solution of Einstein's equations for a vacuum with cosmological constant, which represents a generalization of a flat de Sitter solution to the Bianchi I anisotropic case. In a particular case, the author showed that the plane-symmetric de Sitter universe develops from a cigar singularity of Kasner type towards an asymptotically isotropic de Sitter universe. Our solutions are completely different from those of Gron since we have considered non-empty universes without cosmological constant. However, the late time behavior of our model coincides with that of Gron, as the solutions in both models approach a de Sitter stage.

The models for $\alpha\neq0$ have been divided into four cases, i.e., (i) $0<\alpha<1$, (ii) $1<\alpha<3$, (iii) $\alpha>3$, and (iv) $\alpha=3$. The solutions in the first three cases have also been extended to the (normal or phantom) scalar field model to know the behavior of effective matter. The first case corresponds to an accelerating universe, while the other three cases represent decelerating universes. If we respect the NEC and WEC, then the model with $0<\alpha<1$ describes only late time acceleration. The scalar field model shows that late time acceleration takes place in the presence of quintessence DE. The model can also describe a early inflationary phase in the presence of a phantom scalar field, but at a cost of violation of the NEC and WEC.

The solutions with $1<\alpha<3$ hold NEC and WEC at late times and describe decelerated universes. The scalar field model shows that decelerated phases occur in the presence of dust and/or radiation. In case of $\alpha>3$, the NEC and WEC are obeyed during early evolution, but the DEC is always violated. The violation of the DEC also reflects in the scalar field model where the scalar potential becomes negative. Hence, the solutions in this case are physically unrealistic. Nevertheless, if one accepts a negative scalar potential then the solutions can also be consistent with quintessence scalar field. In the last case ($\alpha=3$), the solutions are mathematically different from those of $\alpha\neq3$. The model in this particular case exhibits a stiff matter era.

The models in all these cases, except $\alpha\geq3$, have also been found to be anisotropic and have shear during the early evolution, but become isotropic and shear free at late times. The stiff matter phase ($\alpha=3$) has shear at early stages and becomes shear free at late times, but it evolves with constant anisotropy which is insignificant. However, the ``cosmic no hair conjecture" is not obeyed in any of these cases, which is obvious for power law expansion.

Finally, we would like to mention that all the above mentioned outcomes will also be valid in a general Bianchi I space-time model.

\end{document}